\documentclass[aps,prl,twocolumn]{revtex4-2}
\usepackage[utf8]{inputenc} 
\usepackage[T1]{fontenc}    
\usepackage{hyperref}       
\usepackage{url}            
\usepackage{booktabs}       
\usepackage{amsfonts}       
\usepackage{microtype}      
\usepackage{xcolor}         
\usepackage{braket}         
\usepackage{varioref}
\usepackage{hyperref}
\usepackage{amsmath} 
\usepackage{nameref}        
\usepackage{graphicx}       
\usepackage[ruled,vlined]{algorithm2e}  
\usepackage{enumitem}       
\usepackage{amsthm}         
\usepackage{xspace}         
\usepackage{subfig}         
\usepackage{bbold}          
\usepackage{qcircuit}       
\usepackage{float}          
\usepackage{svg}            
\usepackage{array}
\usepackage{mathtools}

\usepackage{chngcntr}

\newcounter{mysfig}
\counterwithin{mysfig}{figure}

\renewcommand\themysfig{\thefigure(\alph{mysfig})}
\makeatletter
\newcommand\Scaption[1]{%
\refstepcounter{mysfig}%
\vskip.5\abovecaptionskip
  \sbox\@tempboxa{\small\themysfig~#1}%
  \ifdim \wd\@tempboxa >\hsize
    \small\themysfig~#1\par
  \else
    \global \@minipagefalse
    \hb@xt@\hsize{\hfil\box\@tempboxa\hfil}%
  \fi
  \vskip\belowcaptionskip}
\makeatother

\usepackage{todonotes}
\usepackage{comment}

\hypersetup{
    colorlinks=true,
    urlcolor=cyan,
}

\newtheoremstyle{cited}%
  {3pt}
  {3pt}
  {\itshape}
  {}
  {\bfseries}
  {.}
  {.5em}
  {\thmname{#1} \thmnumber{#2} \thmnote{\normalfont#3}}

\theoremstyle{cited}

\theoremstyle{plain}
\newtheorem{lemma}{Lemma}

\theoremstyle{definition}
\newtheorem{definition}{Definition}
\newtheorem{citedth}{Theorem}
\newtheorem{citedcoro}{Corollary}

\newcommand{\R}{\mathbb{R}}

\renewcommand{\1}{\mathbb{1}} 

\newcommand{\cH}{\mathcal{H}}

\newcommand{\cN}{\mathcal{N}}

\newcommand{\cM}{\mathcal{M}}
\newcommand{\cL}{\mathcal{L}}

\newcommand{\mat}[1]{\left(\begin{matrix}#1\end{matrix}\right)}

\SetKwComment{Comment}{$\vartriangleright$}{}

\begin{document}
\title{Optimality conditions for spatial search with multiple marked vertices}

\author{Mathieu Roget}
\author{Hachem Kadri}
\author{Giuseppe Di Molfetta}
\email{giuseppe.dimolfetta@lis-lab.fr}
\affiliation{Aix Marseille Univ,  Université de Toulon, CNRS, LIS, Marseille, France}

\begin{abstract}

We contribute to fulfil the long-lasting gap in the understanding of the spatial search with multiple marked vertices. The theoretical framework is that of discrete-time quantum walks (QW), \textit{i.e.} local unitary matrices that drive the evolution of a single particle on the lattice. QW based search algorithms are well understood when they have to tackle the fundamental problem of finding only one marked element in a $d-$dimensional grid and it has been proven they provide a quadratic advantage over classical searching protocols. However, once we consider to search more than one element, the behaviour of the algorithm may be affected by the spatial configuration of the marked elements and even the quantum advantage is no longer guaranteed. Here our main contribution is threefold~: (i)~we provide \textit{sufficient conditions for optimality} for a multi-items QWSearch algorithm~; (ii)~we provide analytical evidences that \textit{almost, but not all} spatial configurations with multiple marked elements are optimal; and (iii)~we numerically show that the computational advantage with respect to the classical counterpart is not always certain and it does depend on the proportion of searched elements over the total number of grid points. 
\end{abstract}

\maketitle

One of the main applications of quantum computing is algorithmic. Considered to be still beyond the reach of today's quantum computers, it has had a major impact in several fields, from cryptography~\cite{gisin2002quantum} and quantum machine learning~\cite{biamonte2017quantum} to quantum simulation~\cite{georgescu2014quantum}. The first quantum algorithms were formulated in the early 1990s~\cite{deutsch1992rapid, shor1994algorithms}, and since then researchers have continued to create new ones over the past 30 years \footnote{For a comprehensive catalog of quantum algorithms visit~\url{https://quantumalgorithmzoo.org}}, trying to optimise computing time and quantum resources. However, compared to the thousands of non-quantum algorithms, the number of quantum algorithms is still modest. This is essentially due to the difficulty of proving the advantage that each of them has over its classical counterpart. An example is given by one of the most studied problem in computer science : \textit{the quantum search} in an unstructured set of $N$ elements. The first algorithm aiming to solve this problem appeared in L. K. Grover~\cite{grover1996fast}'s work. The basic ideas was to introduce a quantum oracle, which recognizes a solution to a search problem when it sees one. The Grover algorithm could solve this problem in $O\left(\sqrt{N}\right)$ time, \textit{i.e.} quadratically faster than what a classical computer needs to complete the same task, and it soon seemed to be extraordinary advantageous to speed-up many classical algorithms that use search heuristic~\cite{bennett1997strengths, ambainis2006quantum, casale2020quantum}. In fact the Grover search algorithm can be applied to any decision problems whose solutions can be checked efficiently~\cite{montanaro2016quantum}, with a clear polynomial speed-up. Such quantum advantage has been proven for several generalizations, for instance when the target elements are multiple~\cite{boyer1998tight}. 
However, the situation may be dramatically different for \textit{spatial} searching, as in~\cite{roget2020grover}, where quadratic speed-up is known to be possible only for some specific case. Spatial search may come in different forms, in continuous time and discrete time. The first example of spatial search algorithm in continuous time has been introduced by Childs and Goldstone~\cite{childs2004spatial} in the quantum walk framework, where the searching method involves now a Hamiltonian defined over an arbitrary graph, which has to be able to solve the searching problem. In this context, it has been proven that only for some certain graphs, such as a complete graph or the hypercube, the hitting time shows a quadratic speed-up with respect to the classical counterpart. This long-standing problem has been recently addressed by Chakraborty, Novo and Roland~\cite{chakraborty2016spatial}, who obtained the necessary and sufficient conditions for the Childs and Goldstone algorithm to be optimal for any graph that meets certain general spectral properties. Another open problem that has remained poorly understood until now is the \textit{spatial search of multiple target items}, both in continuous and discrete time. While the hitting time in the case we search only one target item is in line with the one recovered by the Grover algorithm, when the items are multiple, their spatial configuration can affect significantly the performance (\textit{aka} the time complexity) of the algorithm. The intuitive reason behind it is that the marked vertices interfere among them, and the interplay between constructive and destructive interference may determine very different scenario. Such scenario has been remarked first by Aaronson and Ambainis~\cite{aaronson2003quantum} and recently observed by Bezerra et al~\cite{bezerra2021quantum}, but never been studied and fully understood. Yet traditionally, it has been assumed that when the number of elements searched is low, the spatial search algorithm is optimal. In this paper, we will prove this assumption false. In fact, we will prove that even when considering only two searched elements, there is always a set of spatial configurations for which the search algorithm is \textit{sub-optimal}. Fortunately, this need not worry because this number is small compared to the number of optimal configurations, for all practical purposes. More specifically, we will show that almost all configurations with two marked elements are optimal (complexity in $O(\sqrt N \text{polylog}(N)$) and we will able to precisely upper bound the number of non-optimal configurations (complexity in $O(N)$). As a byproduct of this rigorous result, we will provide a set of \textit{sufficient conditions} on the searched elements' relative position on the grid to ensure the optimality of the algorithm.    
Finally, we discuss the results and we provide strong numerical evidences that such non-optimality issue is not only present when the number of marked elements is two. More in general, the quantum advantage will be shown to be not always guaranteed and also strictly depending on the ratio of marked elements $M/N$, with $M$ the number of the searched elements on the grid.

%
\paragraph{A 2-dimensionnal, discrete time Quantum Walk Search (QWSearch) algorithm}
A discrete time quantum walk on a grid is the quantum analog of a 2-dimensional random walk. The quantum walker lives in a composite Hilbert space : the coin state space, encoding the walker direction and the position state space. The physical space here is a grid of size $\sqrt{N}\times \sqrt{N}$. A generic state of the walker reads as follows:
\begin{small}
\begin{equation}\nonumber
\ket{\psi} = \sum_{v\in \{0,1\} } \sum_{x=0}^{\sqrt{N-1}}\sum_{y=0}^{\sqrt{N-1}} \alpha_{v,x,y} \ket{v,x,y},
\end{equation}
\end{small}
where the coin state space is spanned by the $z$-basis $\{\ket{0},\ket{1}\}$. The walker evolves driven by the usual split-step operator~:
\begin{equation}\nonumber
    U = \Sigma_y (C_y\otimes \mathbb{I}_N) \Sigma_x (C_x\otimes \mathbb{I}_N),
\end{equation}
where $C_y$ and $C_x$ are two non commutative $U(2)$ operators and the $\Sigma_i$, are coin state dependent shift operator along direction $i = x,y$, defined as follows~: 
\begin{align}\nonumber
    \Sigma_i \ket{v}\ket{i} &= \ket{v}\ket{i-(-1)^v}.
\end{align}
The search algorithm based on the aforementioned quantum walk scheme (QWSearch), for a set of marked vertices $|m\rangle\in \cM$ is implemented as follows: (i) Initialize the quantum walker $\psi(0)$ to the equal superposition over all states, by applying $n$ single bit Hadamard operation to the ground state $\ket{0}$. (ii) Given a coin oracle $R = \1 - 2\sum_{m\in \cM}\ket{d,m}\bra{d,m}$, where $\ket{d}$ is the diagonal state in the coin state space, apply the perturbed evolution operator $U' = UR$, for $t_{opt}$ steps, the hitting time. (iii) Measure the state in the $\ket{v,x,y}$ basis. The success probability of the walker $p(t)$ after $t$ time steps is given by :
\begin{small}
\begin{multline}\nonumber\label{eq:proba_succ}
    p(t) = \sum_{m\in \cM}|\bra{d,m}{U'}^t\ket{\psi(0)}|^2 =\\
    \sum_{j=0,1}|\sum_{\theta}e^{i \theta t}\braket{d,m_j|\theta}\braket{\theta|\psi(0)}|^2.
\end{multline}
\end{small}
where the operator $U'$ has been diagonalised on its basis. The main aim here is to find the hitting time $t_{opt}$ which maximises the above probability. Assuming that the search algorithm converges in finite time, one shall expect that the eigenspace of $U'$ be approximately spanned by $\{\ket{\lambda_+},\ket{\lambda_-}\}$, with $e^{i\lambda_+}$ and $e^{i\lambda_-}$ the two closest eigenvalues to unity~: 
\begin{multline}\nonumber
\lambda_+ = \min_{e^{i\theta}\in Sp(U'),\; \theta >0}\theta \;\text{ and }\; \lambda_- = \max_{e^{i\theta}\in Sp(U'),\; \theta <0}\theta.
\end{multline}
Thus, let us cast all the negligible contributions as $\epsilon_m$ :  
\begin{small}
\begin{equation}\label{eq:proba_succ1}
p(t) = \sum_{j=0,1}|\beta_{+,j}e^{i\lambda_+ t}+\beta_{-,j}e^{i\lambda_- t}+ \epsilon_m|^2.
\end{equation}
\end{small}
To analytically calculate the probability of success, we must therefore obtain an explicit expression of the coefficients $\beta_{i,j} = \braket{d,m_j|\lambda_i} \braket{\lambda_i|\psi(0)}$. To begin with,
let us consider the eigenmodes of the QW operator $\ket{\psi_{\pm k,l}}$, with eigenvalues $e^{i \phi_{\pm k,l}}$. We can point out that $\braket{\psi_{\pm k,l}|\lambda_i}$, can be expressed in terms of  $\braket{d,m_j|\lambda_i}$. In fact~:
\begin{small}
\begin{multline}\nonumber
\braket{\psi_{\pm k,l}|U'|\lambda_i} = e^{i \phi_{\pm k,l}}\braket{\psi_{\pm k,l}|\lambda_i}\\
-2e^{i \phi_{\pm k,l}}\sum_{j=0,1}\braket{\psi_{\pm k,l}|d,m_j}\braket{d,m_j|\lambda_i}
\end{multline}
\end{small}
which leads to 
\begin{small}
\begin{multline}
\label{eq:psieq}
    \braket{\psi_{\pm k,l}|\lambda_i} =\\ \frac{2}{1-e^{i (\lambda_i-\phi_{\pm k,l})}}\sum_{j=0,1}\braket{\psi_{\pm k,l}|d,m_j}\braket{d,m_j|\lambda_i}.
\end{multline}
\end{small}
Moreover, since $\lambda_i$ is close to zero, the pre-factor of the sum can be approximated as follows~:
\begin{equation}\nonumber
\frac{2}{1-e^{i (\lambda_i-\phi_{\pm k,l})}} = 1 +i b_{\pm k,l}^\lambda,
\end{equation}
where 
\begin{small}
\begin{equation}\nonumber\label{eq:b_equiv}
b_{\pm,k,l}^{\lambda} = 
\left\{\begin{matrix}
\frac{2}{\lambda_i} + O(\lambda_i)& \text{if }\phi_{\pm,k,l} = 0\\
\frac{-1}{1-\cos\phi_{\pm,k,l}}\left(\lambda_i+\sin\phi_{\pm,k,l}\right) + O(\lambda_i^2) & \text{otherwise}.\\
\end{matrix}\right.
\end{equation}
\end{small}
From Eq.~\ref{eq:psieq} and using the completeness relation, we recover the characteristic equation~: 
\begin{equation}
\label{eq:EqLvec}
\Lambda^\lambda \braket{d,m_j|\lambda_i} = 0,
\end{equation} 
where the symmetric square matrix $\Lambda$ has elements 
\begin{small}
\begin{equation}\nonumber
    \Lambda_{j,j'}^\lambda =  \sum_{a=\pm,k,l} b_{a,k,l}^{\lambda} \braket{d,m_j\mid\psi_{a,k,l}}\braket{\psi_{a,k,l}\mid d,m_{j'}}.
\end{equation}
\end{small}
\
Leaving a detailed calculation to the supplementary material, it can be shown that there exists two distinct cases~:
\vspace{0.2cm}
\paragraph{Case i} For $x+y$ odd, then 
\begin{eqnarray}
\label{eq:Lam0}
 N\Lambda_{j,j'}^\lambda \sim  \left\{\begin{matrix}\frac{4}{\lambda} - \lambda\frac{N \ln{N}}{\pi} & \text{if } j = j'\\ 
0 & \text{otherwise},\\
\end{matrix}\right.
\end{eqnarray}
\paragraph{Case ii} For $x+y$ even~:
\begin{eqnarray}
\label{eq:Lam}
 N\Lambda_{j,j'}^\lambda \sim \left\{\begin{matrix}\frac{4}{\lambda} - \lambda\frac{N \ln{N}}{\pi} & \text{if } j = j'\\ 
\frac{4}{\lambda} - \mathcal{I} - \lambda\mathcal{M} & \text{otherwise},\\
\end{matrix}\right.
\end{eqnarray}
where 
\begin{small}
\begin{align}
\label{eq:caseII}
\mathcal{I} &= \frac{-1}{2}\sum_{k,l=0}^{\sqrt N -1}\cos\left(2\pi\frac{kx+ly}{\sqrt N}\right)\frac{\sin\frac{2\pi k}{\sqrt N}\sin\frac{2\pi l}{\sqrt N}}{1-\cos\frac{2\pi k}{\sqrt N}\cos\frac{2\pi l}{\sqrt N}}.& \nonumber\\
\hspace{0.1cm}
\mathcal{M} &= \sum_{k,l =0}^{\sqrt N -1} \frac{\cos\left(2\pi\frac{kx+ly}{\sqrt N}\right)}{1-\cos\frac{2\pi k}{\sqrt N}\cos\frac{2\pi l}{\sqrt N}}.& 
\end{align}
\end{small}
Now, being the columns of $\Lambda$ linearly dependent, its determinant is zero. Thus, using Eqs.~\ref{eq:Lam0},\ref{eq:Lam}, we can compute
the roots of the equation $\det \Lambda = 0$~ respectively for~:

\vspace{0.2cm}
\textit{Case~i}~:
\begin{equation}\label{eq:lbd1}\nonumber
\lambda_\pm \sim \pm 2 \sqrt{\frac{\pi}{N\ln(N)}},
\end{equation}
and for \textit{Case ii}~:
\begin{equation}\label{eq:lbd2}\nonumber
\lambda_\pm \sim \frac{- \mathcal{I} \pm \sqrt{\mathcal{I}^2 + 32 (\frac{N \ln{N}}{\pi} + \mathcal{M})}}{2 (\frac{N \ln{N}}{\pi} + \mathcal{M})}.
\end{equation}

Having calculated lambdas allows us to explicitly get all the coefficients $\beta_{i,j}$ in the Eq.~\ref{eq:proba_succ1}. Let us study \textit{Case i} and \textit{Case ii} separately. 
In the former, the $\lambda_i$ are equal up to a sign, thus the Eq.~\ref{eq:proba_succ1} reduces to~:
\begin{equation}
\label{eq:reduceP}
p(t) \sim 4 |\beta_{+,j}|^2\sin^2 (\lambda t + \text{cte}) + O(\epsilon_m)
\end{equation}
where $\lambda = |\lambda_\pm|$ and $4 |\beta_{+,j}|^2 \sim \frac{\pi}{8\ln N}$ and the hitting time $t_{opt} \sim \frac{\sqrt{\pi N \ln N}}{4}$. Thus the overall complexity time is $O(\sqrt{N \ln N})$ for a success probability scaling as $O(\ln^{-1} N)$, in line with previous QW search for only one element on the grid. We know that this bound is unlikely to be improved, given the strong arguments given by \cite{magniez2012hitting, patel2010search, santha2008quantum}. This is not unexpected because the QW operator acts independently onto the sub-lattice $\cL_e = \{(i,j) \mid i+j\text{ even}\}$, with even vertices and the one with odd vertices $\cL_o = \{(i,j) \mid i+j\text{ odd}\}$, as we can see in Fig.~\ref{fig:signal}.a. This is due to the bipartite nature of a QW in discrete time makes. So that, if the two searched elements are located respectively on white cells and black cells of a chessboard, the QWsearch is running two parallel and independent searching.
\begin{figure}[h!]
     \centering
     \stepcounter{figure}
    \begin{minipage}[t]{\columnwidth}
    \begin{minipage}[c]{\columnwidth}
    \includegraphics[width=\linewidth]{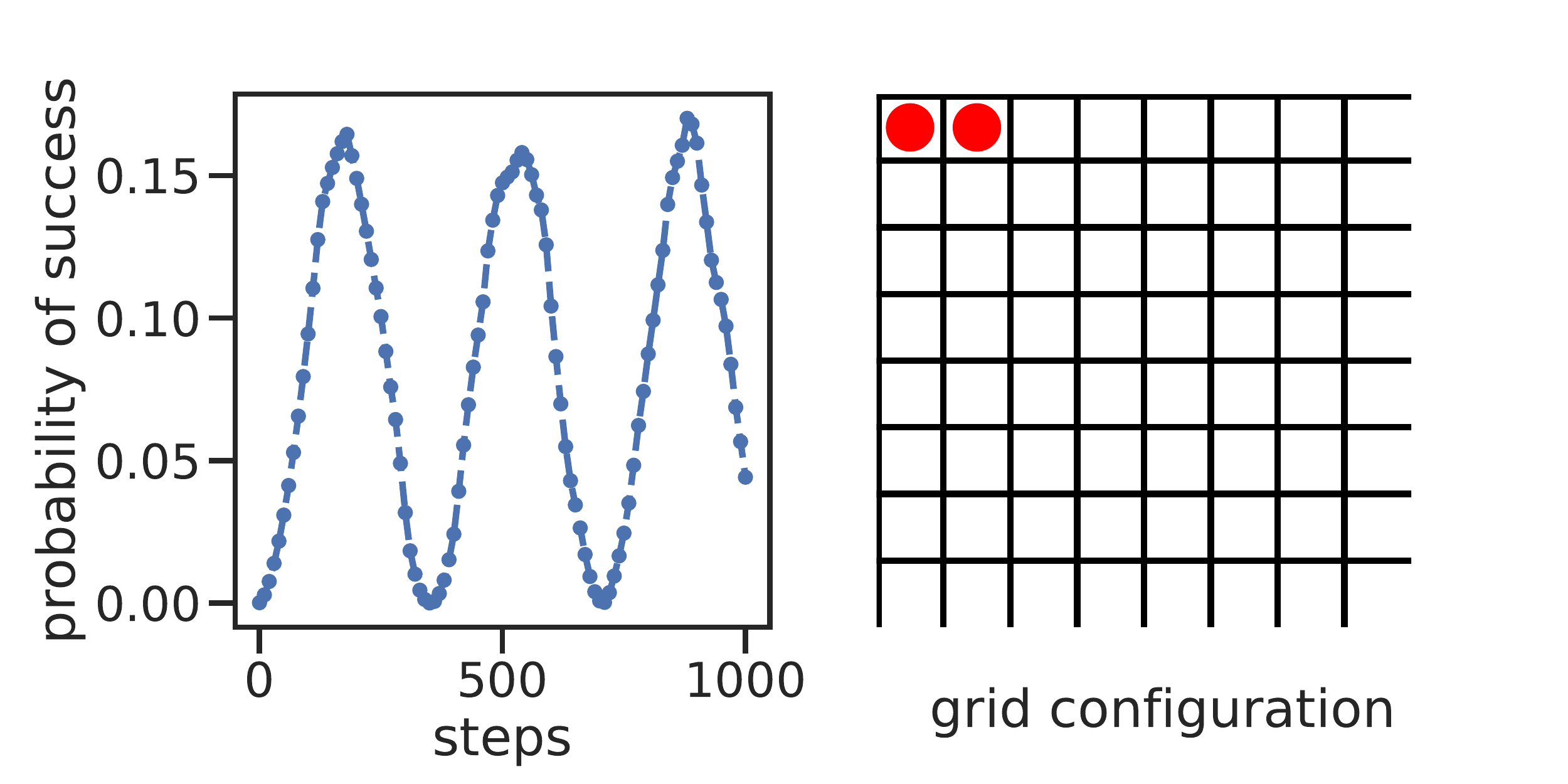}\par
    \end{minipage}
    \Scaption{Optimal configuration. The searched elements are respectively on the odd and even partition of the grid. The two elements are \textit{searched} independently. The algorithm is always optimal for this configuration.}
    \end{minipage}
     \hfill
     \begin{minipage}[t]{\columnwidth}
    \begin{minipage}[b]{\columnwidth}
    \includegraphics[width=\linewidth]{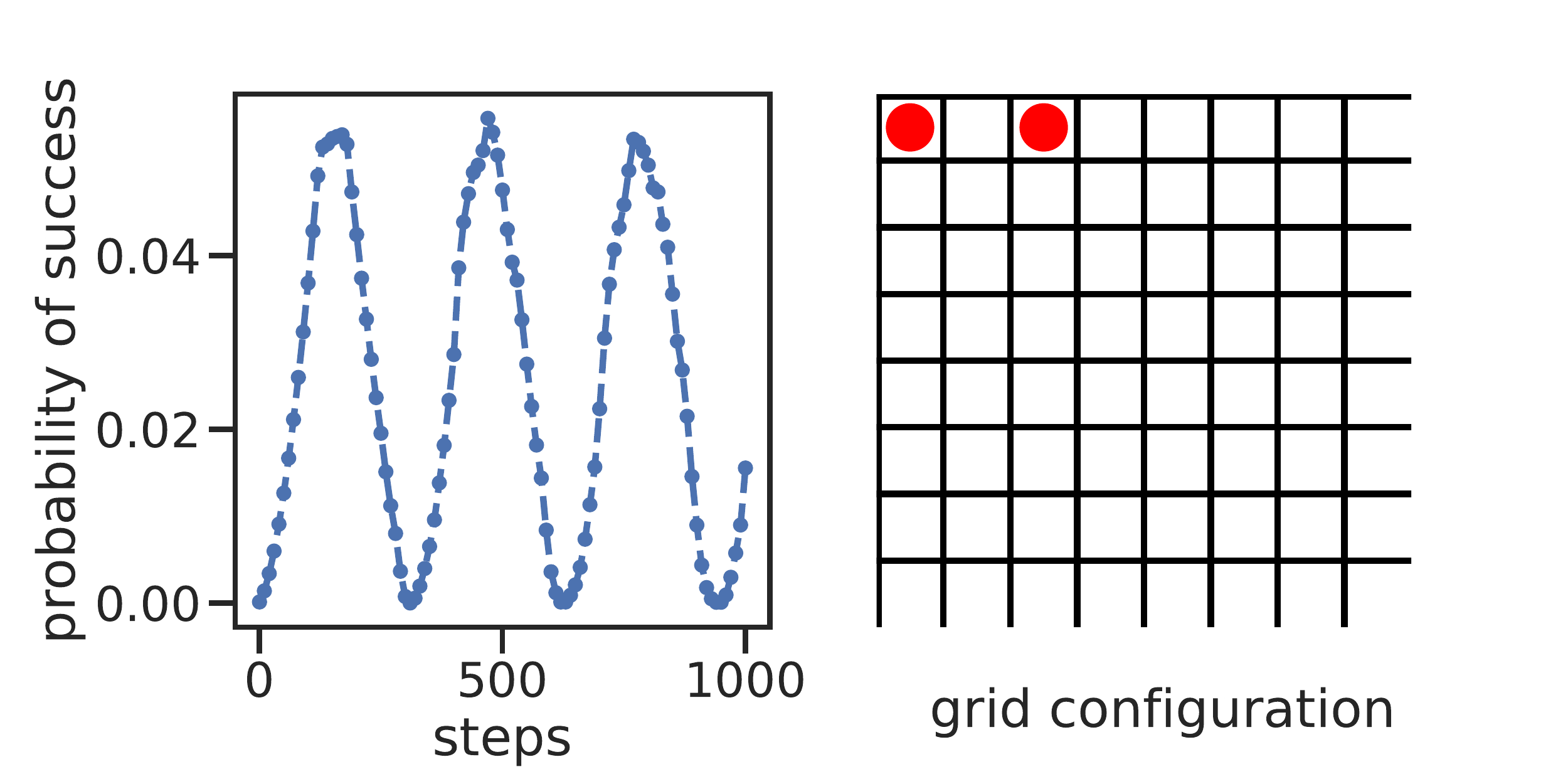}\par
    \end{minipage}
    \Scaption{Optimal configuration. Here the optimality of the algorithm is ensured by the condition $\mathcal{I}\ll \mathcal{M}+\frac{N \ln{N}}{\pi}$, leading to a success probability $O(1/\ln N)$}
    \end{minipage}
    \begin{minipage}[t]{\columnwidth}
    \begin{minipage}[b]{\columnwidth}
    \includegraphics[width=\linewidth]{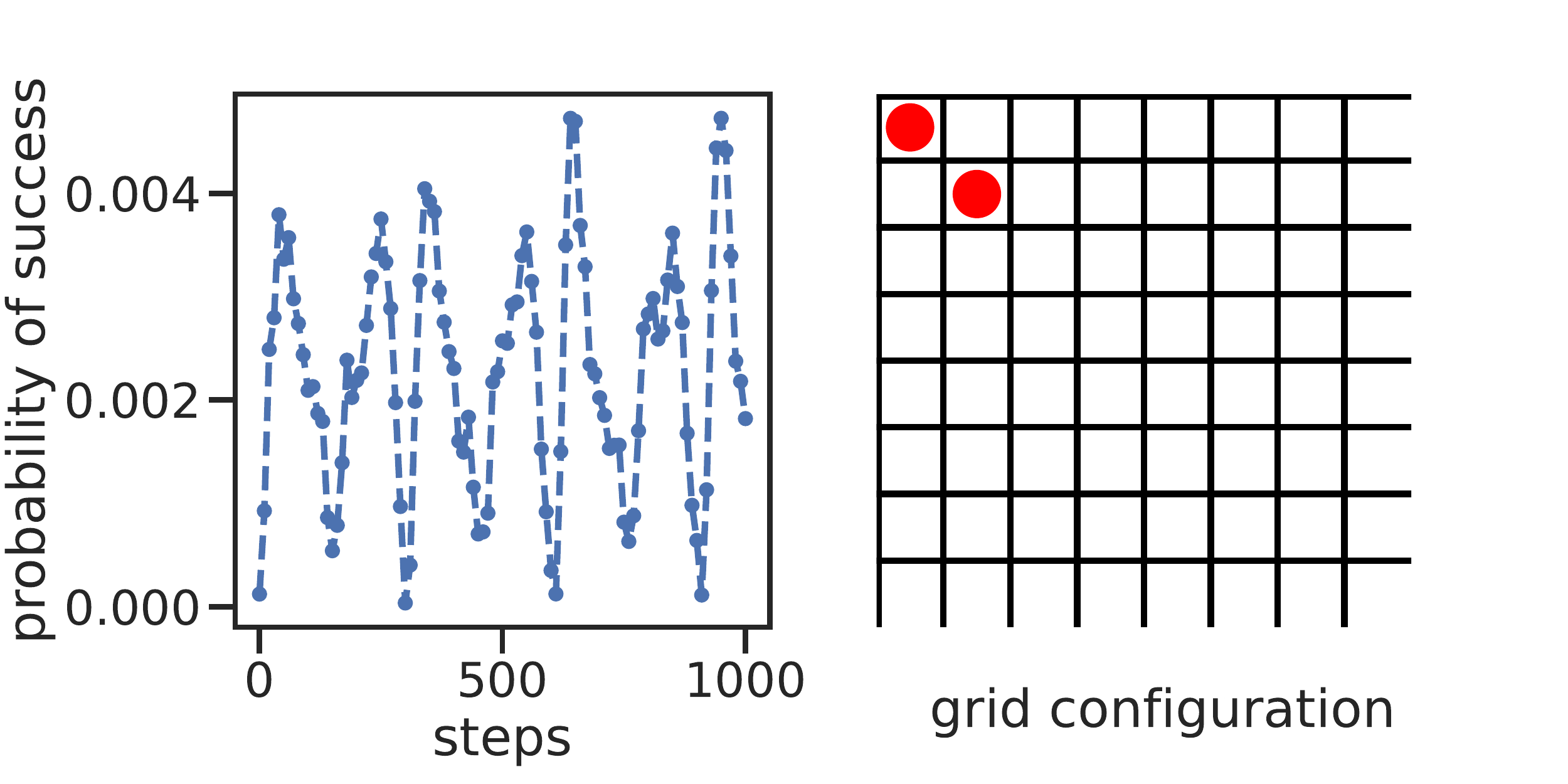}\par
    \end{minipage}
    \Scaption{Non-optimal configuration. For this configuration the condition $\text{min}(x,n-x) \text{min}(y,n-x)\geq n $ does not hold, leading to a success probability $O(1/N)$ }
    \end{minipage}
    \addtocounter{figure}{-1}
    \caption{Three examples of spatial configuration for which the QWSearch algorithms behaves differently.}
    \label{fig:signal}
\end{figure}
Now let us consider \textit{Case ii}. The coefficients $\beta_{i,j}~=~\braket{d, m_j| \lambda_i}\braket{\lambda_i |\psi(0)}$ in the Eq.~\ref{eq:proba_succ1} read~:
\begin{equation*}
    |\beta_{i,j}|^2 \sim \frac{N}{\lambda_i^2(\frac{N\ln N}{\pi}+\mathcal M)^2}.
\end{equation*}
Notice that there are cases where we can manifestly recover optimality. One example is given in Fig.~\ref{fig:signal}.b. Indeed, in the limit
 \begin{equation}
 \label{eq:condition}
 \mathcal{I}^2\ll \mathcal{M}+\frac{N \ln{N}}{\pi}
 \end{equation}
the $\lambda_\pm \sim \pm\frac{2\sqrt{2}}{\sqrt{\mathcal{M}+\frac{N \ln{N}}{\pi}}},
$
and again the success probability reduces to Eq.~\ref{eq:reduceP} 
with $\lambda = |\lambda_\pm|$ and $ |\beta_{+,j}|^2 \sim \frac{N}{8(\frac{N\ln N}{\pi}+\mathcal M)}$. This results lead to the very same complexity we have in  \textit{Case i}. It is easy to convince oneself that condition expressed by the Eq.~\ref{eq:condition} is satisfied in a finite number of cases.  
In fact, after have computed the sums in the Eq.~\ref{eq:caseII} and simplified for large $n = \sqrt{N}$,  
$\mathcal{I} = O\left(\frac{n^2}{\text{min}(x,n-x) \text{min}(y,n-x)}\right)$ and 
$0 \leq \mathcal{M} \leq N \log N$.
Then the sufficient condition for a given spatial configuration to be optimal is  
$\text{min}(x,n-x) \text{min}(y,n-x)\geq n.$
But how many of them do not meet the above condition~? It is possible to prove that they are always bounded by $O(n \ln{n})=O(\sqrt{N} \ln{\sqrt{N}})$. Although finite, this number is statistically negligible for large values of $n$. In fact, if one tosses a random configuration, then the probability to find a non-optimal configuration, such as one in Fig.~\ref{fig:signal}.c, is given by  $O(\sqrt{N}\ln{\sqrt{N}}/N)\xrightarrow{N\rightarrow \infty}0$. \\

\begin{figure}[]
    \centering
    \includegraphics[width=8.6cm]{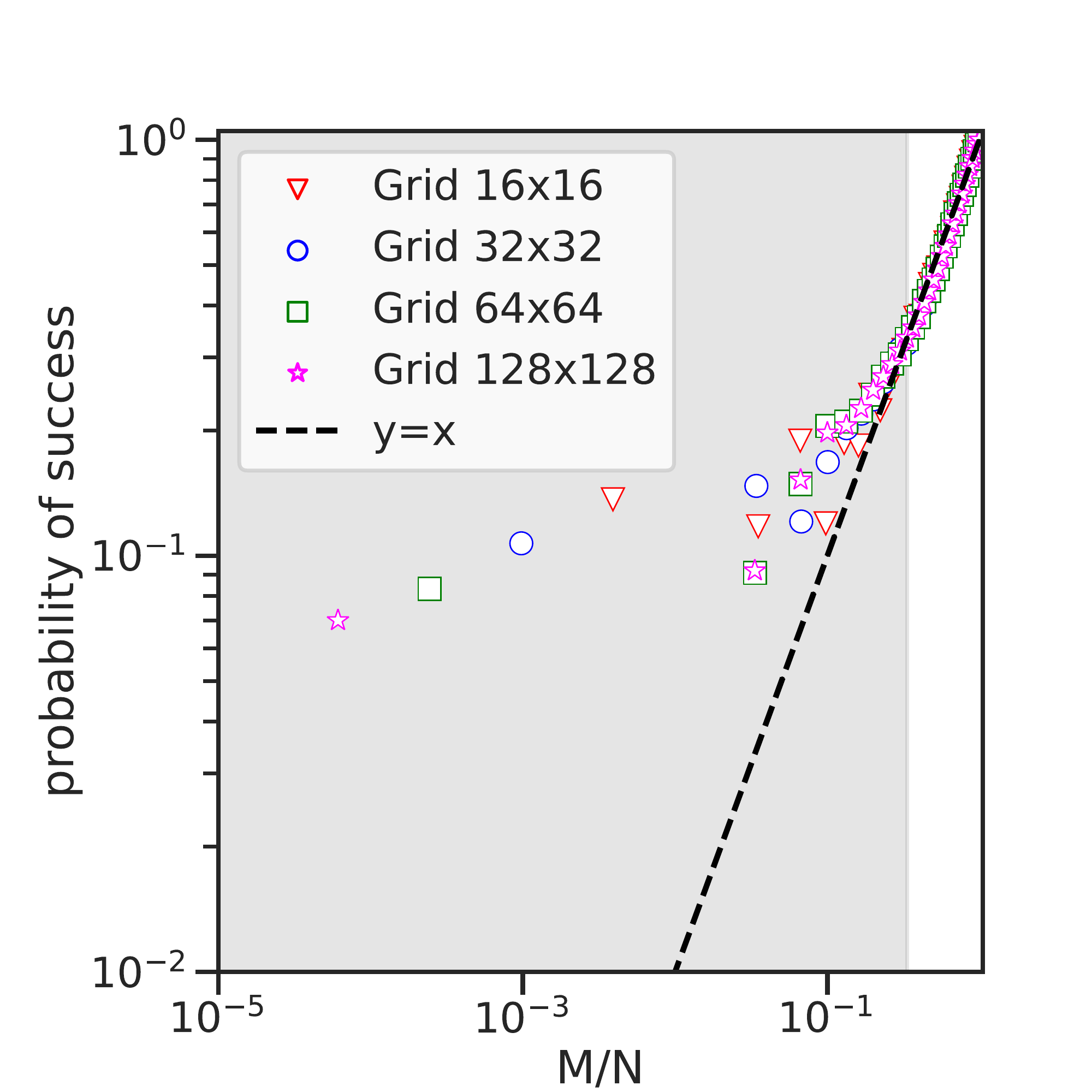}
    \caption{Average probability of success of QW Search assuming uniform distribution of the marked elements configuration in function of $\tau=M/N$.}
    \label{fig:proba_vs_tau}
\end{figure}

Extending the results obtained in this letter to more than two elements is mathematically prohibitive, although feasible in theory. In any case, however, it is possible to show that the number of searched elements, at constant grid size, dramatically affects the advantage over classical search algorithms. Let's note $M$ the number of marked vertices and $\tau = M/N$. In order to search $M$ vertices, we may define the following procedure: (i) Shuffle the grid to get a random configuration; (ii) Apply the searching algorithm for $t_{opt}=\lfloor\sqrt{\pi N \ln N}/4\rfloor$ steps; (iii) Measure the final state over the computational basis of the walker. The Fig.~\ref{fig:proba_vs_tau} shows the average success probability in function of $\tau$ for several grid sizes. Quite remarkably the success probability does not depend on the grid size. Furthermore, for $\tau$ sufficiently large, 
the quantum advantage is completely lost and the success probability coincides with the classical one. 

\begin{figure}[]
    \centering
    \includegraphics[width=8.6cm]{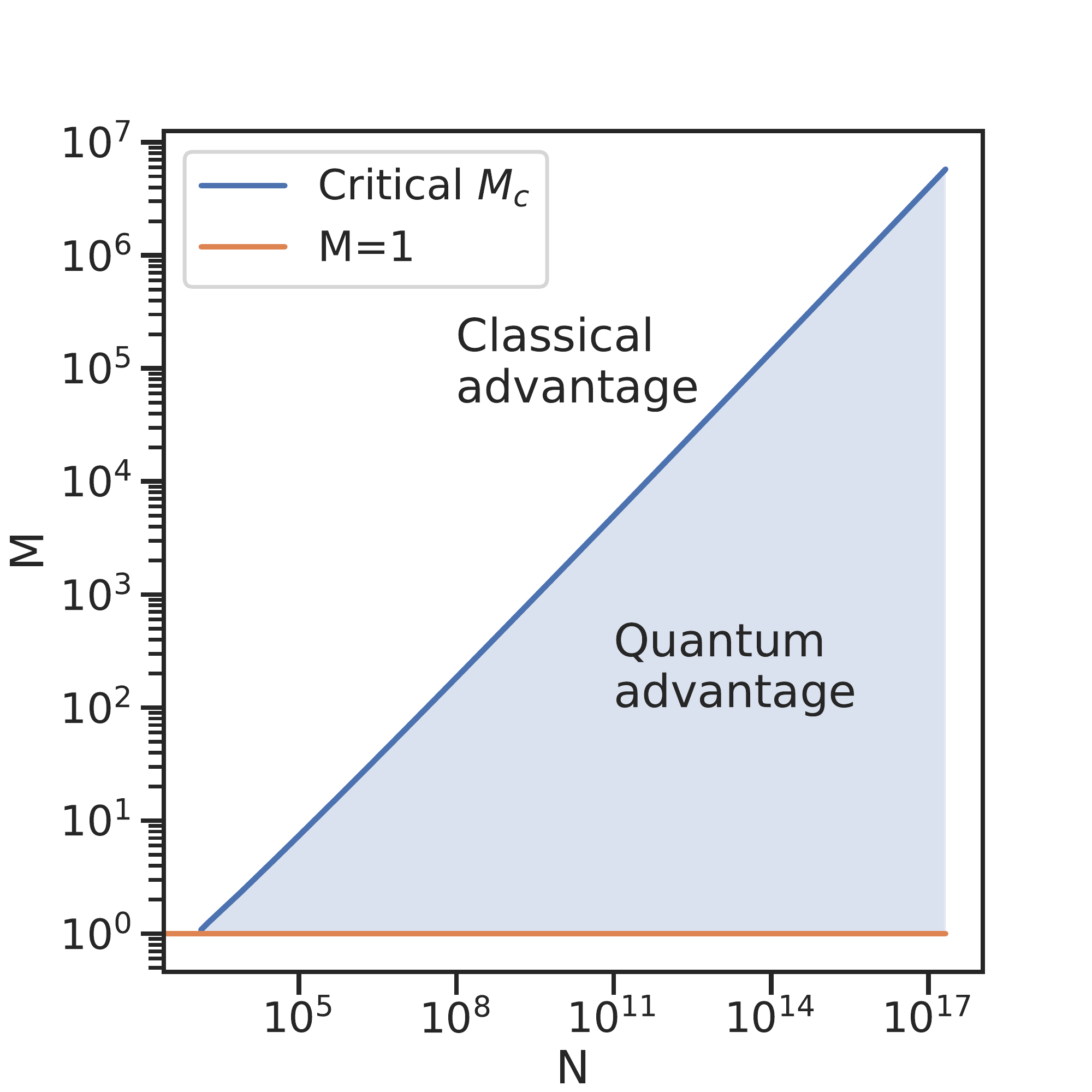}
    \caption{Critical number of marked elements $M_c$ in function of the grid size $N$.}
    \label{fig:M_vs_N}
\end{figure}

To best characterize at what point the advantage is lost, in Fig.~\ref{fig:M_vs_N} we compare the success probability of our quantum algorithm with respect to a classical algorithm. For the same number of query, we consider a stochastic search algorithm with success probability~: $
    p_{cl} = 1 - \frac{\binom{N-t_{opt}}{M}}{\binom{N}{M}} \sim 1-(1-\tau)^{t_{opt}}$.
Using a numerical solver (here a binary search), Fig.~\ref{fig:M_vs_N} shows the critical number of marked vertices $M_c$ upon which the quantum advantage is lost (i.e. the classical algorithm has a higher probability of success). 

%
\paragraph{Discussion}
In this letter, we have provided definitive proof that the spatial search algorithm, based on a quantum discrete-time walker, is not always optimal. We have proved the existence of an upper bound on the number of sub-optimal configurations, which remains small for large grid sizes. The proof was obtained in the special case of only two elements searched on the rectangular grid. We have also shown that there is strong numerical evidence that the advantage of the quantum algorithm depends not only on the spatial configuration but also on the ratio of the number of elements searched over the grid size. Such evidence is crucial in order to be able to use the quantum search algorithm correctly, being sure of a quantum advantage, albeit a polynomial one. The sufficiency conditions for optimality are thus relevant for a wide range of applications where the searching algorithm is used as subroutine~: from simulation to optimisation, from machine learning to distributed algorithmic. 

\paragraph{Acknowledgements} This work is supported by grant DisQC ANR-22-CE47-0002-01 from the French National Research Agency and the Amidex fondation. 

\paragraph{Author Contribution Statement}
The authors confirm contribution to the paper as follows: study conception and design~: GDM and MR; data collection: MR; analysis and interpretation of results: GDM, MR, HK; draft manuscript preparation~: GDM and MR. All authors reviewed the results and approved the final version of the manuscript.

\bibliographystyle{ieeetr} 
\bibliography{biblio}

\onecolumngrid
\appendix
\section{Supplementary material}\label{annex:1}

In this section, we give further detail on the computations ranging from Equation \ref{eq:EqLvec} to the coefficients $\beta_{i,j}$. 

Before that, we provide the explicit expressions for the eigenvalues and eigenvectors of $U$~:
\begin{eqnarray}
    \label{eq:U_explicit}
        \ket{\psi_{a,k,l}}&=&\ket{v_{a,k,l}}\otimes \frac{1}{\sqrt{N}}\sum_{j,j'}\exp\left(2i\pi\frac{kj+lj'}{\sqrt{N}}\right)\ket{j,j'}\\
       \phi_{a,k,l}&=& a \arccos\left(\cos\frac{2\pi k}{\sqrt{N}}\cos\frac{2\pi l}{\sqrt{N}}\right)\\
       |\braket{d\mid v_{a,k,l}}|^2&=& \frac{1}{2} - a\frac{1}{4} \frac{\sin\frac{2\pi k}{\sqrt{N}}\sin\frac{2\pi l}{\sqrt{N}}}{\sin\phi_{+,k,l}}\1_{\phi_{a,k,l}\neq 0}
\end{eqnarray}

\subsection{Computation of $\Lambda^\lambda$}
We recall that 
\begin{equation}\nonumber
\Lambda_{j,j'}^\lambda =  \sum_{a=\pm,k,l} b_{a,k,l}^{\lambda} \braket{d,m_j\mid\psi_{a,k,l}}\braket{\psi_{a,k,l}\mid d,m_{j'}}.
\end{equation}
In order to simplify the coefficients of $\Lambda^\lambda$, we use the explicit form of $\psi_{a,k,l}$ to show that
\begin{align*}
    \braket{d,m_j\mid\psi_{a,k,l}}\braket{\psi_{a,k,l}\mid d,m_{j'}} &= \braket{d\mid v_{a,k,l}}\braket{m_j\mid \psi_{a,k,l}}\braket{v_{a,k,l}\mid d}\braket{\psi_{a,k,l}\mid m_{j'}}\\
    &= \frac{|\braket{d\mid v_{a,k,l}}|^2}{N} \exp\left(2i\pi\frac{k(x_j-x_{j'})+l(y_j-y_{j'})}{\sqrt{N}}\right).
\end{align*}
then,
\begin{equation}\label{eq:Lambda_coef}
\braket{d,m_j\mid\psi_{a,k,l}}\braket{\psi_{a,k,l}\mid d,m_{j'}} = \left\{\begin{matrix}\frac{|\braket{d\mid v_{a,k,l}}|^2}{N} & \text{if } j = j'\\ 
\frac{|\braket{d\mid v_{a,k,l}}|^2}{N} e^{\pm 2i\pi\frac{kx+ly}{\sqrt{N}}} & \text{otherwise}.\\
\end{matrix}\right.
\end{equation}
The main idea now is to use Eq.~\ref{eq:b_equiv} to cut the coefficients $\Lambda^\lambda_{j,j'}$ into three cases. Let us start for the coefficients $\Lambda^\lambda_{j,j}$~:
\begin{align*}
    N\Lambda^\lambda_{j,j} &= N\sum_{a=\pm,k,l} b_{a,k,l}^{\lambda} \braket{d,m_j\mid\psi_{a,k,l}}\braket{\psi_{a,k,l}\mid d,m_{j}} & \text{according to Equation \ref{eq:EqLvec}}\\
    &= \sum_{a=\pm,k,l} b_{a,k,l}^{\lambda} |\braket{d\mid v_{a,k,l}}|^2 & \text{using Equation \ref{eq:Lambda_coef}}\\
    &\sim \sum_{\phi_{a,k,l}=0} \frac{2}{\lambda} |\braket{d\mid v_{a,k,l}}|^2 - \sum_{\phi_{a,k,l}\neq 0} \frac{(\sin \phi_{a,k,l})+\lambda}{1-\cos \phi_{a,k,l}} |\braket{d\mid v_{a,k,l}}|^2 & \text{using Equation \ref{eq:b_equiv}}\\
    &= \frac{1}{\lambda}\sum_{\phi_{a,k,l}=0} 1 + \frac{1}{2}\sum_{\phi_{+,k,l}\neq 0} \frac{\sin \frac{2\pi k}{\sqrt{N}} \sin \frac{2\pi l}{\sqrt{N}}}{1-\cos\frac{2\pi k}{\sqrt{N}}\cos\frac{2\pi l}{\sqrt{N}}} - \lambda \sum_{\phi_{+,k,l}\neq 0}\frac{1}{1-\cos\frac{2\pi k}{\sqrt{N}}\cos\frac{2\pi l}{\sqrt{N}}} & \text{using Equation \ref{eq:U_explicit}}.
\end{align*}
First, it is straightforward to prove that there are four possible cases for $\phi_{a,k,l}=0$, thus $\sum_{\phi_{a,k,l}=0} 1 = 4$. Furthermore, one can show that the second sum $\displaystyle \frac{1}{2}\sum_{\phi_{+,k,l}\neq 0} \frac{\sin \frac{2\pi k}{\sqrt{N}} \sin \frac{2\pi l}{\sqrt{N}}}{1-\cos\frac{2\pi k}{\sqrt{N}}\cos\frac{2\pi l}{\sqrt{N}}} = \frac{1}{2}\sum_{\phi_{k,l}\neq 0}S_{k,l}$ is vanishing, using the symmetry $S_{k,l}~=~-S_{n-k,l}$. Moreover the last sum can be approximated using integral bounding, resulting in $\displaystyle \sum_{\phi_{+,k,l}\neq 0}\frac{1}{1-\cos\frac{2\pi k}{\sqrt{N}}\cos\frac{2\pi l}{\sqrt{N}}} \sim \frac{1}{\pi}N\ln N$. And finally, 
\begin{equation}
    \Lambda^\lambda_{j,j} \sim \frac{4}{\lambda} - \lambda \frac{N\ln N}{\pi}.
\end{equation}
Let us now compute the other half of the $\Lambda^\lambda$'s coefficients. By using a similar procedure~:  
$$
N\Lambda^\lambda_{j,j} \sim \frac{1}{\lambda}\sum_{\phi_{a,k,l}=0} e^{\pm 2i\pi\frac{kx+ly}{\sqrt{N}}} + \frac{1}{2}\sum_{\phi_{+,k,l}\neq 0} \frac{\sin \frac{2\pi k}{\sqrt{N}} \sin \frac{2\pi l}{\sqrt{N}}}{1-\cos\frac{2\pi k}{\sqrt{N}}\cos\frac{2\pi l}{\sqrt{N}}}e^{\pm 2i\pi\frac{kx+ly}{\sqrt{N}}} - \lambda \sum_{\phi_{+,k,l}\neq 0}\frac{1}{1-\cos\frac{2\pi k}{\sqrt{N}}\cos\frac{2\pi l}{\sqrt{N}}}e^{\pm 2i\pi\frac{kx+ly}{\sqrt{N}}}.
$$
Now, we can study separately \textit{case i} and \textit{case ii}.
\paragraph{Case i: $x+y$ odd}
All three above addends are identically vanishing by symmetry. In particular, for each addend $S_{k+\frac{n}{2},l+\frac{n}{2}} = (-1)^{x+y}S_{k,l}$. Consequently,  
\begin{equation}
    \Lambda^\lambda_{j,j'} = 0,
\end{equation}
with $j\neq j'$ and $x+y$ odd.
\paragraph{Case ii: $x+y$ even}
The first sum $\sum_{\phi_{a,k,l}=0} e^{\pm 2i\pi\frac{kx+ly}{\sqrt{N}}}=4$. The other two sums are harder to compute. We can, however, retrieve the expressions of $\mathcal{I}$ and $\mathcal{M}$ by using the symmetry $S_{n-k,n-l}=\overline{S_{k,l}}$. This symmetry implies that the sum is real. We can thus discard the imaginary part. In conclusion
$$
N\Lambda^\lambda_{j,j} \sim \frac{4}{\lambda} - \mathcal I - \lambda \mathcal{M},
$$
where 
$$
\mathcal I = \frac{-1}{2}\sum_{\phi_{+,k,l}\neq 0} \frac{\sin \frac{2\pi k}{\sqrt{N}} \sin \frac{2\pi l}{\sqrt{N}}}{1-\cos\frac{2\pi k}{\sqrt{N}}\cos\frac{2\pi l}{\sqrt{N}}}\cos\left(2\pi\frac{kx+ly}{\sqrt{N}}\right) \qquad \text{and} \qquad \mathcal M = \sum_{\phi_{+,k,l}\neq 0}\frac{1}{1-\cos\frac{2\pi k}{\sqrt{N}}\cos\frac{2\pi l}{\sqrt{N}}}\cos\left(2\pi\frac{kx+ly}{\sqrt{N}}\right).
$$

\subsection{Computation of $\lambda_i$ and $|\beta_{i,j}|^2$}
In this section we give further details about how we obtained each $\lambda_i$. We assume that $x+y\text{ even } \Rightarrow \mathcal{I}^2\ll \mathcal{M}+\frac{N \ln{N}}{\pi} $.
We start by using Eqs.~\ref{eq:Lam0} and \ref{eq:Lam} combined with $\det \Lambda^\lambda=0$ (which can be deduced from Eq.~\ref{eq:EqLvec}) to get $\lambda_{\pm}$. Once done, we note that $|\braket{d,m_0\mid\lambda_\pm}^2 = |\braket{d,m_1\mid\lambda_\pm}^2$ because of the symmetry of $\Lambda^\lambda$. Combining the latter with the Eq.~\ref{eq:psieq}, we can deduce that 
\begin{equation}
    |\braket{\lambda_a |\psi(0)}|^2 = \left|1+ib^{\lambda_a}_{+,0,0}\right|^2 \left|\sum_{j=0,1}\braket{\psi_{\pm k,l}|d,m_j}\braket{d,m_j|\lambda_a}\right|^2 \sim \frac{8|\braket{d,m\mid \lambda}|^2}{N\lambda^2}.
\end{equation}
where $\ket{\psi(0)} = \ket{\psi_{+,0,0}}$.
One can similarly use Eq.~\ref{eq:psieq} in $\sum_{\pm,k,l}|\braket{\psi_{\pm,k,l}\mid\lambda}|^2 = 1$ to show that
\begin{equation}\label{eq:coef1}
    |\braket{d,m\mid\lambda}|^2 \sim \frac{1}{4}\frac{N}{\frac{N\ln N}{\pi}+\mathcal M \1_{\{x+y\text{ even}\}}}.
\end{equation}
Combining the above result into the expression of  $|\braket{\lambda_a |\psi(0)}|^2$, we get
\begin{equation}\label{eq:coef2}
    |\braket{\lambda_a |\psi(0)}|^2 \sim \frac{2+2.\1_{x+y\text{ even}}}{\lambda^2\left(\frac{N\ln N}{\pi}+\mathcal M \1_{\{x+y\text{ even}\}}\right)}.
\end{equation}
Again, let us consider \textit{case i} and \textit{case ii} separately.~\\

\paragraph{Case i: $x+y$ odd}
\begin{equation}
    \det \Lambda^\lambda = 0 \Leftrightarrow \frac{4}{\lambda} - \lambda \frac{N\ln N}{\pi} \sim 0 
    \Leftrightarrow \lambda^2\sim 4 \frac{\pi}{N\ln N} \Leftrightarrow \lambda_\pm \sim \pm 2\sqrt{\frac{\pi}{N\ln N}} .
\end{equation}
We recognise the Eq.~\ref{eq:lbd1}. We can now solve explicitly Eqs.~\ref{eq:coef1} and \ref{eq:coef2} for \textit{Case i}~:
\begin{equation}
    |\braket{d,m\mid\lambda_\pm}|^2 \sim \frac{\pi}{4\ln N} \qquad \text{and} \qquad |\braket{\lambda_\pm |\psi(0)}|^2 \sim \frac{1}{2}.
\end{equation}
Notice that the latter expression implies that $\epsilon_m \to 0$. We recover the right expression $4 |\beta_{+,j}|^2 \sim \frac{\pi}{8\ln N}$.~\\

\paragraph{Case ii: $x+y$ even}
\begin{align*}
    \det \Lambda^\lambda = 0 &\Leftrightarrow \left(\Lambda^\lambda_{0,0}\right)^2 - \left(\Lambda^\lambda_{1,0}\right)^2 = 0 & \text{ because $\Lambda^\lambda$ is symmetric}\\
    &\Leftrightarrow \left(\Lambda^\lambda_{0,0} - \Lambda^\lambda_{1,0}\right)\left(\Lambda^\lambda_{0,0} + \Lambda^\lambda_{1,0}\right)=0\\
    &\Leftrightarrow \left[\mathcal I - \lambda\left(\frac{N\ln N}{\pi}- \mathcal M\right)\right]\left[\frac{8}{\lambda}-\mathcal I - \lambda\left(\frac{N\ln N}{\pi}+ \mathcal M\right)\right]=0 & \text{using Equation \ref{eq:Lam}}\\
    &\Leftrightarrow \lambda_\pm \in \left\{ \frac{- \mathcal{I} + \sqrt{\mathcal{I}^2 + 32 (\frac{N \ln{N}}{\pi} + \mathcal{M})}}{2 (\frac{N \ln{N}}{\pi} + \mathcal{M})}, \frac{- \mathcal{I} - \sqrt{\mathcal{I}^2 + 32 (\frac{N \ln{N}}{\pi} + \mathcal{M})}}{2 (\frac{N \ln{N}}{\pi} + \mathcal{M})}, \frac{\mathcal I}{\frac{N \ln{N}}{\pi}-\mathcal M} \right\}.\\
\end{align*}
Using the hypothesis $\mathcal{I}^2\ll \mathcal{M}+\frac{N \ln{N}}{\pi}$ and the fact that $0\leq \mathcal M \leq \frac{N\ln N}{\pi}$, it holds~:
$$
\lambda_{\pm} \sim \frac{- \mathcal{I} - \pm \sqrt{\mathcal{I}^2 + 32 (\frac{N \ln{N}}{\pi} + \mathcal{M})}}{2 (\frac{N \ln{N}}{\pi} + \mathcal{M})} \sim \pm \frac{2\sqrt{2}}{\sqrt{\frac{N \ln{N}}{\pi} + \mathcal{M}}} = O\left(\frac{1}{\sqrt{N\ln N}}\right).
$$
We recognise Eq.~\ref{eq:lbd2}.Again we solve the Eqs.~\ref{eq:coef1} and \ref{eq:coef2} for \textit{Case ii} and note that 
\begin{equation}
    |\braket{d,m\mid\lambda_\pm}|^2 \sim \frac{1}{4}\frac{N}{\frac{N\ln N}{\pi}+\mathcal M} \qquad \text{and} \qquad |\braket{\lambda_\pm |\psi(0)}|^2 \sim \frac{1}{2}.
\end{equation}
As for \textit{case i}, the latter implies $\epsilon_m\to 0$, meaning that all other eigenvalues of $U'$ can be neglected~(including the last solution of $\det \Lambda^\lambda = 0$). We get finally $4 |\beta_{+,j}|^2 \sim \frac{1}{2}\frac{N}{\frac{N\ln N}{\pi}+\mathcal M}$ and using $M \leq N\ln N$, we can deduce that the success probability $p_{succ}~  |\beta_{+,j}|^2 = O(\ln^{-1} N)$.

\end{document}